\documentclass[aps,twocolumn,showpacs,preprintnumbers,amsmath,amssymb,superscriptaddress,floatfix,nofootinbib]{revtex4}

\usepackage{graphicx}
\usepackage{epsfig}
\usepackage{epstopdf}
\usepackage{hyperref}
\usepackage{amsmath}
\usepackage{amsfonts}
\usepackage{amssymb}
\usepackage{color}

\begin{document}

\title{The $a_0(980)$ and $\Lambda(1670)$ in the $\Lambda^+_c \to \pi^+ \eta \Lambda$ decay}

\author{Ju-Jun Xie}
\affiliation{Institute of Modern Physics, Chinese Academy of
Sciences, Lanzhou 730000, China} \affiliation{State Key Laboratory
of Theoretical Physics, Institute of Theoretical Physics, Chinese
Academy of Sciences, Beijing 100190, China}

\author{Li-Sheng Geng} \email{lisheng.geng@buaa.edu.cn}
\affiliation{School of Physics and Nuclear Energy Engineering and
International Research Center for Nuclei and Particles in the
Cosmos, Beihang University, Beijing 100191, China}
\affiliation{State Key Laboratory of Theoretical Physics, Institute
of Theoretical Physics, Chinese Academy of Sciences, Beijing 100190,
China}

\date{\today}

\begin{abstract}

We propose to study the $a_0(980)$ and the $\Lambda(1670)$ resonances in the
$\Lambda^+_c \to \pi^+ \eta \Lambda$ decay via the final state
interactions of the  $\pi^+ \eta$ and $\eta \Lambda$ pairs. The weak
interaction part proceeds through the $c$ quark decay process:
$c(ud) \to (s + u + \bar d)(ud)$, while the hadronization part takes
place in two different mechanisms. In the first mechanism, the $sud$
cluster picks up a $q\bar{q}$ pair from the vacuum to form the
$\eta\Lambda$ meson-baryon pair while the $u\bar{d}$ pair from the
weak decay hadronizes into a $\pi^+$.  In the second, the $sud$
cluster turns into a $\Lambda$, while the $u\bar{d}$ pair from the
$c$ decay picks up a $q\bar{q}$ pair and hadronizes into a
meson-meson pair ($\pi\eta$ or  $K\bar{K}$). Because the final
$\pi^+ \eta$ and $\eta \Lambda$ states are in pure isospin $I = 1$
and $I=0$ combinations, the $\Lambda^+_c \to \pi^+ \eta \Lambda$
decay can be an ideal process to study the $a_0(980)$ and
$\Lambda(1670)$ resonances. Describing the final state interaction
in  the chiral unitary approach, we find that the $\pi^+ \eta$ and
$\eta \Lambda$ invariant mass distributions, up to an arbitrary
normalization, show clear cusp and peak structures, which can be
associated with the $a_0(980)$ and $\Lambda(1670)$ resonances,
respectively. The proposed mechanism can provide valuable information on the nature of these resonances and
can in principle  be test by facilities such as BEPCII.

\end{abstract}


\maketitle

\section{Introduction}

Understanding the nature of  mesons and baryons has always been 
one of the most challenging topics in hadron physics. The new
observations~\cite{Choi:2003ue,Acosta:2003zx,Abazov:2004kp,Ablikim:2013mio,Liu:2013dau,Aaij:2015tga,D0:2016mwd}
have challenged the conventional wisdom that mesons are made of
quark-antiquark pairs and baryons are composed of three quarks. In
this respect, it is not surprising that both the $a_0(980)$ state and the $\Lambda(1670)$ have
generated a lot of interests in their true nature.  The $a_0(980)$ has been suggested of  being either a $q \bar
q$, a tetraquark state, a meson-meson molecule, a glueball, or a
dynamically generated state~\cite{Klempt:2007cp}. Similarly,  the $\Lambda(1670)$ has been found to
be consistent with both a naive three quark picture and
a molecular picture
 dynamically generated from $\eta\Lambda$ and $\bar{K}N$ interactions.

In the chiral unitary approach, the $a_0(980)$ state was shown to be
dynamically generated from the interaction of $\bar K K$ and $\pi
\eta$ treated as coupled channels in isospin $I =
1$~\cite{Oller:1997ti,Nieves:1998hp}. It decays into $\pi \eta$ in
$s$-wave with a total decay width around $170$ MeV. The pole of the
$a_0(980)$ is much tied to the coupled channels and it disappears if
either the $\pi \eta$ or the $K\bar{K}$ channel is discarded.  The
closeness of its pole to the $K\bar{K}$ threshold and its strong
coupling to $K\bar{K}$ have led to the suggestion that the
$a_0(980)$ state might be a cusp
effect~\cite{Oller:1997ti,Nieves:1998hp}. As a result, the
$a_0(980)$ total decay width increases very fast as its mass
increases. For instance, Ref.~\cite{Janssen:1994wn} has claimed a large width for the $a_0(980)$ around $200$ MeV, where the $a_0(980)$
state was studied within a realistic meson-exchange model for the
$\pi \pi$ and $\pi \eta$ interactions.

On the other hand, in Ref.~\cite{Wolkanowski:2015lsa}, it was
claimed that both the $a_0(980)$ and the $a_0(1450)$ resonance emerge
from a single $q \bar{q}$ seed state. This state interacts with
other mesons, giving rise to meson-meson ($MM$) loop contributions
to the corresponding mass. These contributions shift the pole of the
seed state to higher energies, turning into the $a_0(1450)$. For the
$a_0(980)$, the meson cloud eats up the original seed, becoming the
largest component~\cite{Wolkanowski:2015lsa} (see also
Refs.~\cite{vanBeveren:1986ea,Tornqvist:1995ay,Fariborz:2009cq,Fariborz:2009wf}).
By now it is commonly accepted that the $a_0(980)$ is not a standard
$q \bar q$ state but an extraordinary state~\cite{jaffe}.

For the $\Lambda(1670)$ hyperon resonance, based on the analysis of
the available high precision data on $K^- p \to \eta \Lambda$
reaction, it was argued to be a three-quark
state~\cite{Starostin:2001zz}. Such a conclusion is also supported
in the  study of the $K^- p \to \pi^0 \Sigma^0$
reaction at low energies within the chiral quark model~\cite{Zhong:2008km} . However, in
the chiral unitary approach, the $\Lambda(1670)$ resonance can
be dynamically generated from the $s$-wave meson baryon interaction
in the strangeness $S=-1$ sector~\cite{Oset:2001cn}. Experimentally,
the $\Lambda(1670)$ resonance is known to have a strong coupling to the $\eta
\Lambda$ channel~\cite{Agashe:2014kda}. Hence, it is expected that
the $\eta \Lambda$ production is dominated by formation of the
intermediate $\Lambda(1670)$ resonance.

In recent years, it has been shown that the selection of particular modes is
possible in the nonleptonic weak decays of heavy hadrons~\cite{Oset:2016lyh}. In particular, the nonleptonic weak decays of charmed baryons 
can be  useful tools to study hadronic resonances, some of
which are subjects of intense debate about their
nature~\cite{Klempt:2007cp,Crede:2008vw,Chen:2016qju}. In
addition, those weak decays are also helpful to investigate final
state interactions and hence have the potential to bring further
light into  the nature of some puzzling hadrons~\cite{Oset:2016lyh}.
For instance, the $\Lambda^0_b \to J/\psi \Lambda(1405)$ decay was
studied in Refs.~\cite{Roca:2015tea,Feijoo:2015cca}, where the
$\Lambda(1405)$ state is generated in the final state interaction of
the ground state pseudoscalar mesons and octet baryons ($MB$). In Ref.~\cite{Miyahara:2015cja}, the
work of Refs.~\cite{Roca:2015tea,Feijoo:2015cca} on the $\Lambda^0_b
\to J/\psi MB$ weak decays was extended to the $\Lambda^+_c \to
\pi^+ MB$ weak decays. It is shown there that these weak decays
might be ideal processes to study the $\Lambda(1405)$ and
$\Lambda(1670)$ resonances, because they are dominated by the $I =
0$ contribution. In Ref.~\cite{Hyodo:2011js}, the $\pi \Sigma$ mass
distribution was  studied in the $\Lambda^+_c \to \pi^+ \pi \Sigma$
decays with the aim of extracting the $\pi \Sigma$ scattering lengths.
In a recent work~\cite{Lu:2016ogy} the role of the exclusive
$\Lambda^+_c$ decays into a neutron in testing the flavor symmetry
and final state interaction was investigated. It was shown that the
three body nonleptonic decays are of great interest to explore the
final state interaction in $\Lambda_c^+$ decays.

Along this line, in the present work, we revisit the $\Lambda^+_c
\to \pi^+ \eta \Lambda$ decay taking into account not only the $\eta
\Lambda$ final-state-interaction (FSI), but also the FSI of $\pi^+
\eta$, which gives the line shape of the $a_0(980)$ state. The pure
$I = 1$ nature of the $\pi^+ \eta$ channel is particularly
beneficial to the study of the $a_0(980)$ state.

This article is organized as follows. In Sec.~\ref{Sec:Formalism},
we present the theoretical formalism of the weak
$\Lambda^+_c \to \pi^+ \eta \Lambda$ decay, explaining in detail the hadronization and final state interactions of
the $\eta\Lambda$ and $\pi^+\eta$ pairs. Numerical results and discussions are
presented in Sec.~\ref{Sec:Results}, followed by a short summary in the
last section.

\section{Formalism} \label{Sec:Formalism}

\begin{figure}[htbp]
\begin{center}
\includegraphics[scale=1.]{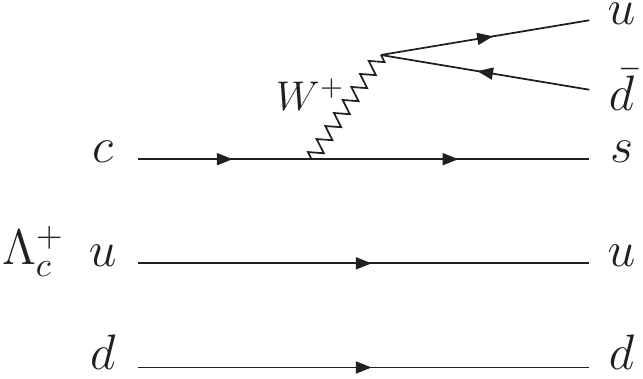}
\caption{The dominant diagram at the quark level for $\Lambda^+_c$
decaying into a $u \bar d$ pair and a $s u d$ cluster. The solid lines and
the wiggly line stand the quarks and the $W^+$ boson, respectively.}
\label{Fig:feynd}
\end{center}
\end{figure}

As shown in Ref.~\cite{Miyahara:2015cja}, a way
for the $\Lambda^+_c \to \pi^+ \eta \Lambda$ to proceed is the
following: in the first step the charmed quark in $\Lambda^+_c$
turns into a strange quark with a $u \bar d$ pair by the weak
decay shown in Fig.~\ref{Fig:feynd}. The next step consists
in introducing a new $\bar q q$ pair with the quantum numbers of
the vacuum, $\bar u u + \bar d d + \bar s s$, to form a meson (baryon) $M (B)$ from
the $s$ quark ($ud$ diquark) or $MM$ from the $u \bar d$ pair with
$\Lambda$ from the $sud$ cluster. Finally, the
final-state interactions (FSIs) of the $MM$ or $MB$ will lead to dynamical generation of the
$a_0(980)$ and $\Lambda(1670)$.  In the following, we discuss the $\eta \Lambda$ and
$\pi^+ \eta$ FSIs separately.

\subsection{Final state interaction of meson-baryon}

\begin{figure}[htbp]
\begin{center}
\includegraphics[scale=1.]{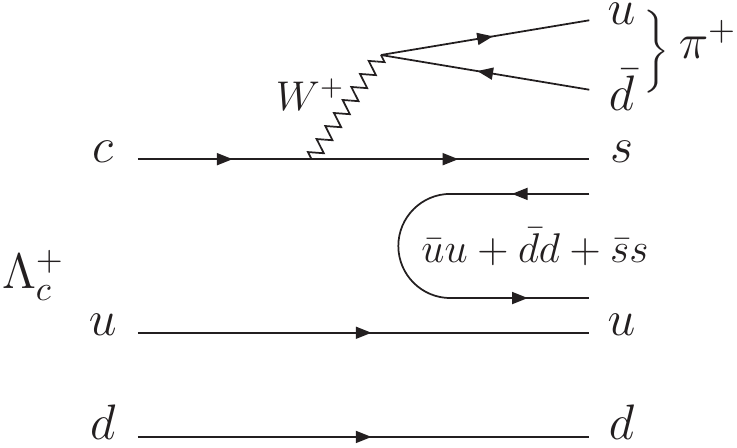}
\caption{Hadronization of the $s u d$ cluster into one meson and one
baryon for the $\Lambda^+_c \to \pi^+ MB$ decay with the $\pi^+$
emission from the $u \bar d$ pair.} \label{Fig:mbproduction}
\end{center}
\end{figure}

We first discuss the decay of $\Lambda^+_c$ to produce the $\pi^+$
from the $u \bar d$ pair and the $sud$ cluster
hadronization into a meson-baryon pair. To create the $MB$ final
state, we must proceed to hadronize the $sud$ cluster by creating an
extra $\bar q q$ pair as depicted in Fig.~\ref{Fig:mbproduction}. In
this process, the $ud$ diquark in $\Lambda^+_c$ is the spectator,
and the $sud$ cluster is combined into a pure $I = 0$ state
\begin{eqnarray}
\frac{1}{\sqrt{2}} |s(ud -du)\rangle.
\end{eqnarray}
As in Refs.~\cite{Roca:2015tea,Miyahara:2015cja}, one can straightforwardly
obtain the meson-baryon states after the $\bar q q$ pair production
as~

\begin{eqnarray}
|MB\rangle = |K^-p\rangle + |\bar{K}^0 n\rangle- \frac{\sqrt{2}}{3} |\eta \Lambda\rangle,
\label{Eq:mbproduction}
\end{eqnarray}
where the flavor states of the baryons are as follows:
\begin{eqnarray}
|p\rangle &=& \frac{1}{\sqrt{2}} |u(ud - du)\rangle, \\
|n\rangle &=&  \frac{1}{\sqrt{2}} |d(ud - du)\rangle, \\
|\Lambda\rangle &=& \frac{1}{\sqrt{12}} |(usd - dsu) + (dsu - uds)
\nonumber \\
&& ~~~~~~~~ + 2(sud - sdu)\rangle.
\end{eqnarray}

\begin{figure}[htbp]
\begin{center}
\includegraphics[scale=0.6]{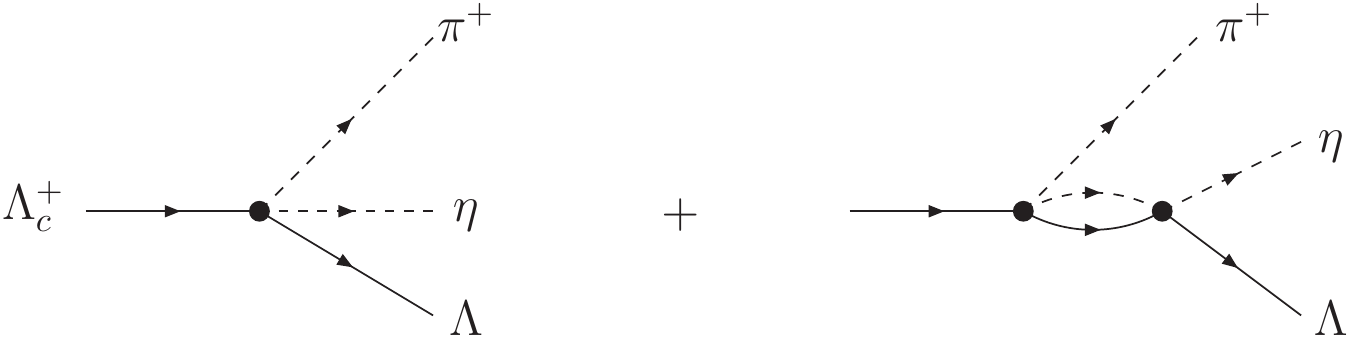}
\caption{The diagram for the meson-meson final-state interaction for
the $\Lambda^+_c \to \pi^+ \eta \Lambda$ decay.} \label{Fig:mbfsi}
\end{center}
\end{figure}

After the production of a meson-baryon pair, the final-state
interaction between the meson and the baryon takes place, which can be
parameterized by the re-scattering shown in Fig.~\ref{Fig:mbfsi}
at the hadronic level for the$\Lambda^+_c \to \pi^+ \eta
\Lambda$  decay. In Fig.~\ref{Fig:mbfsi}, we also
show the tree level diagram for the $\Lambda^+_c \to \pi^+ \eta
\Lambda$ decay.

According to Eq.~\eqref{Eq:mbproduction}, we can write down the
$\Lambda^+_c \to \pi^+ \eta \Lambda$ decay amplitude of
Fig.~\ref{Fig:mbfsi} as,
\begin{eqnarray}
T^{MB} &=& V_P \Big( -\frac{\sqrt{2}}{3} + G_{K^- p} (M_{\eta
\Lambda}) t_{K^- p \to \eta \Lambda} (M_{\eta \Lambda}) \nonumber \\
&& + G_{\bar{K}^0 n} (M_{\eta \Lambda}) t_{\bar{K}^0 n \to \eta
\Lambda} (M_{\eta \Lambda}) \nonumber \\
&& -\frac{\sqrt{2}}{3} G_{\eta \Lambda} (M_{\eta \Lambda}) t_{\eta
\Lambda \to \eta \Lambda} (M_{\eta \Lambda}) \Big),   \label{Eq:tmb}
\end{eqnarray}
where $V_P$ expresses the weak and hadronization strength, which is
assumed to be independent of the final state interaction. In the
above equation, $G_{MB}$ denotes the one-meson-one-baryon loop
function, which depends on the invariant mass of the final $\eta
\Lambda$ system, $M_{\eta \Lambda}$. The meson-baryon scattering
amplitudes $t_{MB \to \eta \Lambda}$ are those obtained in the chiral
unitary approach, which depend also on $M_{\eta \Lambda}$. Details
can be found in Refs.~\cite{Oset:2001cn,Oller:2000fj}.

\subsection{Final state interaction of meson-meson}

\begin{figure}[htbp]
\begin{center}
\includegraphics[scale=1.]{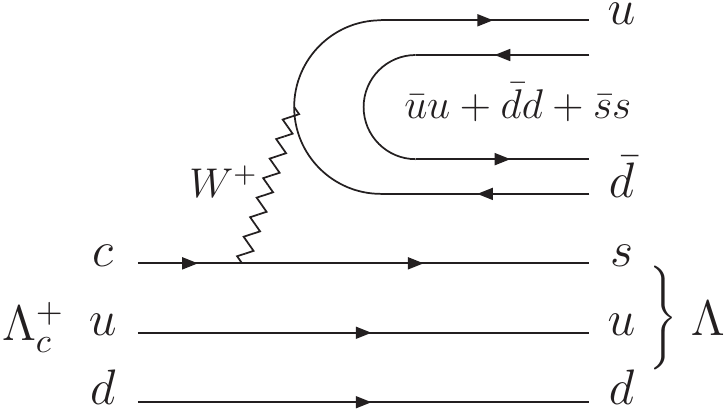}
\caption{The hadronization of the $u \bar d$ pair into two mesons.}
\label{Fig:mmproduction}
\end{center}
\end{figure}

Next, we discuss the decay of $\Lambda^+_c$ to produce the $\Lambda$
from the $sud$ cluster and the $u \bar d$ pair undergoes
hadronization to form a meson-meson pair. The hadronization is now
realized combining an extra $\bar q q$ pair from the vacuum with the $u \bar d$ pair,
as shown in Fig.~\ref{Fig:mmproduction}. In this process, the $I =
0$ $sud$ cluster will form the $\Lambda$ state as
\begin{eqnarray}
\frac{1}{\sqrt{2}} |s(ud -du)\rangle ~~ \Rightarrow ~~ \frac{\sqrt{6}}{3}
|\Lambda\rangle. \label{Eq:sudtoLambda}
\end{eqnarray}

In the following, we explain in detail  which mesons are produced in
the hadronization of the $u \bar d$ pair. We first introduce the $q \bar
q$ matrix
\begin{eqnarray}
M = \left(
           \begin{array}{ccc}
             u\bar u & u \bar d & u\bar s \\
             d\bar u & d\bar d & d\bar s \\
             s\bar u & s\bar d & s\bar s \\
           \end{array}
         \right) = \left(
           \begin{array}{c}
            u   \\
             d  \\
             s   \\
         \end{array}
         \right) \left(
           \begin{array}{ccc}
            \bar{u} & \bar{d} & \bar{s}
           \end{array}   \right),
\end{eqnarray}
which has the followng property
\begin{eqnarray}
M \cdot M = M \times (\bar u u + \bar d d + \bar s s).
\end{eqnarray}

Next, we rewrite the $\bar q q$ matrix $M$ in terms of meson
components,  and as a result $M$ can be identified with the matrix
$\phi$~\cite{Bramon:1992kr,Roca:2003uk,Gamermann:2008jh}
\begin{equation}\label{Eq:phimatrix}
\renewcommand{\arraystretch}{1.5}
\phi = \left(
           \begin{array}{ccc}
             \frac{\eta}{\sqrt{3}} + \frac{\pi^0}{\sqrt{2}} +  \frac{\eta'}{\sqrt{6}} & \pi^+ & K^{+} \\
             \pi^- &\frac{\eta}{\sqrt{3}} - \frac{\pi^0}{\sqrt{2}} +  \frac{\eta'}{\sqrt{6}}  & K^{0} \\
            K^{-} & \bar{K}^{0} & - \frac{\eta}{\sqrt{3}} + \sqrt{\frac{2}{3}}\eta' \\
          \end{array}
         \right),
\end{equation}
which incorporates the standard $\eta$ and $\eta'$
mixing~\cite{Bramon:1992kr}.

\begin{figure}[htbp]
\begin{center}
\includegraphics[scale=0.6]{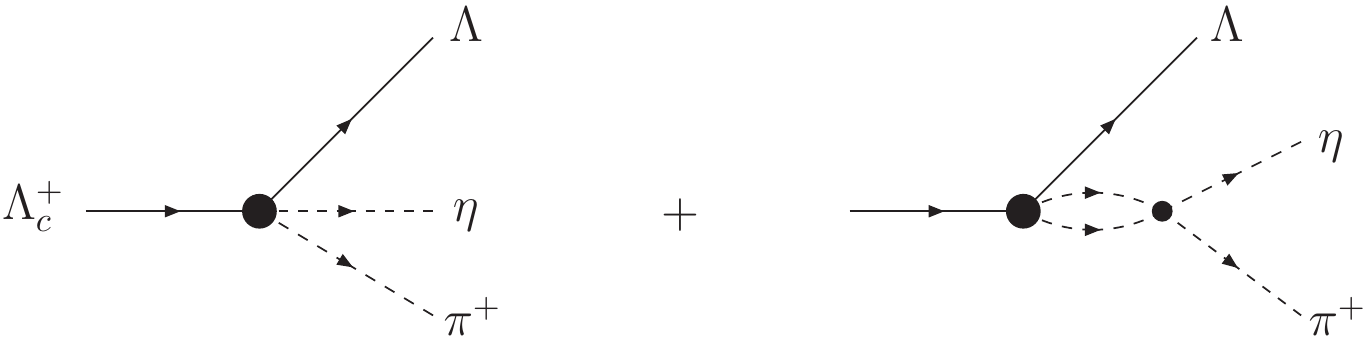}
\caption{The diagram for the meson-meson final-state interaction for
the $\Lambda^+_c \to \pi^+ \eta \Lambda$ decay.} \label{Fig:mmfsi}
\end{center}
\end{figure}

Then, in terms of mesons, the hadronized $u \bar d$ pair is
given by
\begin{eqnarray}
u \bar d (\bar u u +\bar d d +\bar s s) & \equiv & (M \cdot M)_{12} \equiv (\phi \cdot \phi)_{12} \nonumber\\
 & = & \frac{2}{\sqrt{3}} \pi^+ \eta + K^+ \bar K^{0},
\label{Eq:phiphi12}
\end{eqnarray}
where we have omitted the $\eta'$ term because of its large mass.
Taking into account the $\Lambda$ production of Fig.~\ref{Fig:mmfsi} and
the weight of Eq.~\eqref{Eq:sudtoLambda}, we obtain the meson-meson production with a $\Lambda$  baryon  as
\begin{eqnarray}
|MM\rangle = \frac{2\sqrt{2}}{3} \pi^+ \eta + \frac{\sqrt{6}}{3} K^+
\bar{K}^0,
\end{eqnarray}
where the re-scattering of $\pi^+ \eta$ and $K^+ \bar{K}^0$ will
give the line shape of the $a_0(980)$ state.

More specifically the transition amplitude is of the following form:
\begin{eqnarray}
T^{MM} &=& V'_P \Big( \frac{2\sqrt{2}}{3} + \frac{2\sqrt{2}}{3}
G_{\pi^+ \eta} (M_{\pi^+ \eta}) t_{\pi^+ \eta \to \pi^+ \eta}
(M_{\pi^+ \eta})
\nonumber \\
&&  + \frac{\sqrt{6}}{3}G_{K^+ \bar{K}^0} (M_{\pi^+ \eta}) t_{K^+
\bar{K}^0 \to \pi^+ \eta} (M_{\pi^+ \eta})  \Big), \label{Eq:tmm}
\end{eqnarray}
where $G_{M M}$ is the loop function of the two intermediate meson
propagators~\cite{Oller:1997ti} and $t_{MM \to \pi^+ \eta}$ are the
meson-meson scattering amplitudes obtained in
Ref.~\cite{Oller:1997ti}, which depend on the invariant mass
$M_{\pi^+ \eta}$ of the $\pi^+ \eta$ system. In general, the factor
$V'_P$ should be different from the factor $V_P$ in
Eq.~\eqref{Eq:tmb}.

\subsection{Invariant mass distributions of the $\Lambda^+_c \to \pi^+ \eta \Lambda$ decay}

With all the ingredients obtained in the previous section, one can
write down the invariant mass distributions for the $\Lambda^+_c \to
\pi^+ \eta \Lambda$ decay as~\cite{Agashe:2014kda}:
\begin{eqnarray}
\frac{d^2 \Gamma_{\Lambda^+_c \to \pi^+ \eta \Lambda}}{d M^2_{\pi^+
\eta} d M^2_{\eta \Lambda}} &=& \frac{1}{(2\pi)^3}
\frac{M_{\Lambda}}{8 M^2_{\Lambda^+_c}} \left\lvert T (M_{\pi^+
\eta}, M_{\eta \Lambda}) \right\rvert^2 , \label{Eq:dgdmdm}
\end{eqnarray}
where $T$ is the total decay amplitude for the decay of $\Lambda^+_c
\to \pi^+ \eta \Lambda$ depending on the invariant $\pi^+ \eta$ and
$\eta \Lambda$ masses. The $\Lambda_c^+$ decay is a three body decay
and the invariant mass distribution with respect to any of the two
invariant masses is evaluated by integrating over the other
invariant mass. For a given value of $M_{\pi^+ \eta}$, the range of
$M_{\eta \Lambda}$ is fixed as,
\begin{eqnarray}
M^{\rm max}_{\eta \Lambda} \! &=& \! \sqrt{ \left(E_{\eta} +
E_{\Lambda} \right)^2-\left(\sqrt{E^{2}_{\eta} - m^2_{\eta}} - \sqrt{E^{2}_{\Lambda} - m^2_{\Lambda}} \right)^2 }, \nonumber \\
M^{\rm min}_{\eta \Lambda} \! &=& \! \sqrt{ \left(E_{\eta} +
E_{\Lambda} \right)^2 - \left(\sqrt{E^{2}_{\eta} - m^2_{\eta}} +
\sqrt{E^{2}_{\Lambda} - m^2_{\Lambda}} \right)^2 }, \nonumber
\end{eqnarray}
where $E_{\eta} = (M^2_{\pi^+ \eta} - m^2_{\pi^+} +
m^2_{\eta}/2M_{\pi^+ \eta}$ and $E_{\Lambda} = (M^2_{\Lambda^+_c} -
M^2_{\pi^+ \eta} - m^2_{\Lambda})/2M_{\pi^+ \eta}$ are the energies
of $\eta$ and $\Lambda$ in the $\pi^+ \eta$ rest frame. Similarly,
for a given value of $M_{\eta \Lambda}$, we can easily obtain the
range of invariant masses allowed for the $\pi^+\eta$, namely  $M_{\pi^+ \eta}$.

Taking $M_{\pi^+ \eta} = 980$ MeV, we obtain,
\begin{eqnarray}
M^{\rm max}_{\eta \Lambda} &=& 2043 ~{\rm MeV}, \nonumber \\
M^{\rm min}_{\eta \Lambda} &=& 1680 ~{\rm MeV}, \nonumber
\end{eqnarray}
which indicates that the meson-meson final state interaction to form
the $a_0(980)$ state only contributes to $\eta \Lambda$ invariant
masses beyond the peak around $1670$ MeV.
Noting the fact that the obtained peak of the $\Lambda(1670)$ resonance is
narrow~\cite{Oset:2001cn}, we expect that the meson-meson
final state interaction will not affect much the peak structure of the
$\Lambda(1670)$  resonance in the $\eta \Lambda$  invariant  mass
distribution.

Similarly, if we take $M_{\eta \Lambda} = 1670$ MeV,  we obtain,
\begin{eqnarray}
M^{\rm max}_{\pi^+ \eta} &=& 1104 ~{\rm MeV}, \nonumber \\
M^{\rm min}_{\pi^+ \eta} &=& 1011 ~{\rm MeV}, \nonumber
\end{eqnarray}
which seems to imply that the meson-baryon final state interaction to
form the $\Lambda(1670)$ state contributes to the $\pi^+ \eta$ invariant
masses beyond the peak/cusp in the $\pi^+ \eta$ mass around $980$
MeV of the $a_0(980)$ state. However, we will see that the
meson-baryon final state interaction does affect the $a_0(980)$
structure in the invariant $\pi^+ \eta$ mass distribution.

Because the factors $V_P$ and $V'_P$ are unknown in our model, and
the relative phase between $T^{MB}$ and $T^{MM}$ is unknown either,
we will study three models: Model A takes into account only the
meson-baryon final state interaction; Model B includes only the
meson-meson final state interaction; Model C considers both
meson-baryon and meson-meson final state interactions. Because the
relative strong phase, $\delta$, between these two decay mechanisms
is unknown, in Model C, we take $\delta$ as a free parameter. For
these three Models, we can write the total decay amplitude $T$ as,
\begin{eqnarray}
T &=& T^{MB}, ~~~ {\rm for~ Model~ A}, \\
T &=& T^{MM}, ~~~ {\rm for~ Model~ B}, \\
T &=& T^{MB} + \left|\frac{T^{MM}}{T^{MB}}\right|T^{MB}e^{i\delta},
~~~ {\rm for~ Model~ C},
\end{eqnarray}
where the relative strong phase $\delta$ is defined as,
\begin{eqnarray}
\frac{T^{MM}}{T^{MB}} = \left|\frac{T^{MM}}{T^{MB}}\right| e^{i
\delta}.
\end{eqnarray}

On the other hand, since the values of $V_P$ and $V'_P$ are unknown,
we impose a constraint  on the values of $V_P$ and $V'_P$ such that
\begin{eqnarray}
\Gamma^{\rm Model ~A}_{\Lambda^+_c \to \pi^+ \eta \Lambda} =
\Gamma^{\rm Model ~B}_{\Lambda^+_c \to \pi^+ \eta \Lambda}.
\label{Eq:constraint}
\end{eqnarray}
In this respect, we assume that the strengths of the two decay
mechanisms shown in Figs.~\ref{Fig:mbproduction} and
\ref{Fig:mmproduction} are equal. This is a reasonable assumption, since any proposed mechanism should first explain the
experimental decay width, and then the corresponding invariant mass distributions will allow one to distinguish between different decay mechanisms. At present, neither the decay width
nor the invariant mass distribution of the $\Lambda_c\rightarrow\pi^+\eta\Lambda$ is known experimentally. 
From Eq.~\eqref{Eq:constraint}, we
get~\footnote{Since we take also the relative strong phase between
$T^{MB}$ and $T^{MM}$ into account, we discard the other
solution, $V'_P = -0.38 V_P$. Furthermore, in the following
calculations, we assume $V_P$ to be constant and take $V_P = 1$
MeV$^{-1}$.}
\begin{eqnarray}
V'_P = 0.38 V_P.
\end{eqnarray}

\section{Numerical results and discussion} \label{Sec:Results}

\begin{figure}[htbp]
\begin{center}
\includegraphics[scale=0.45]{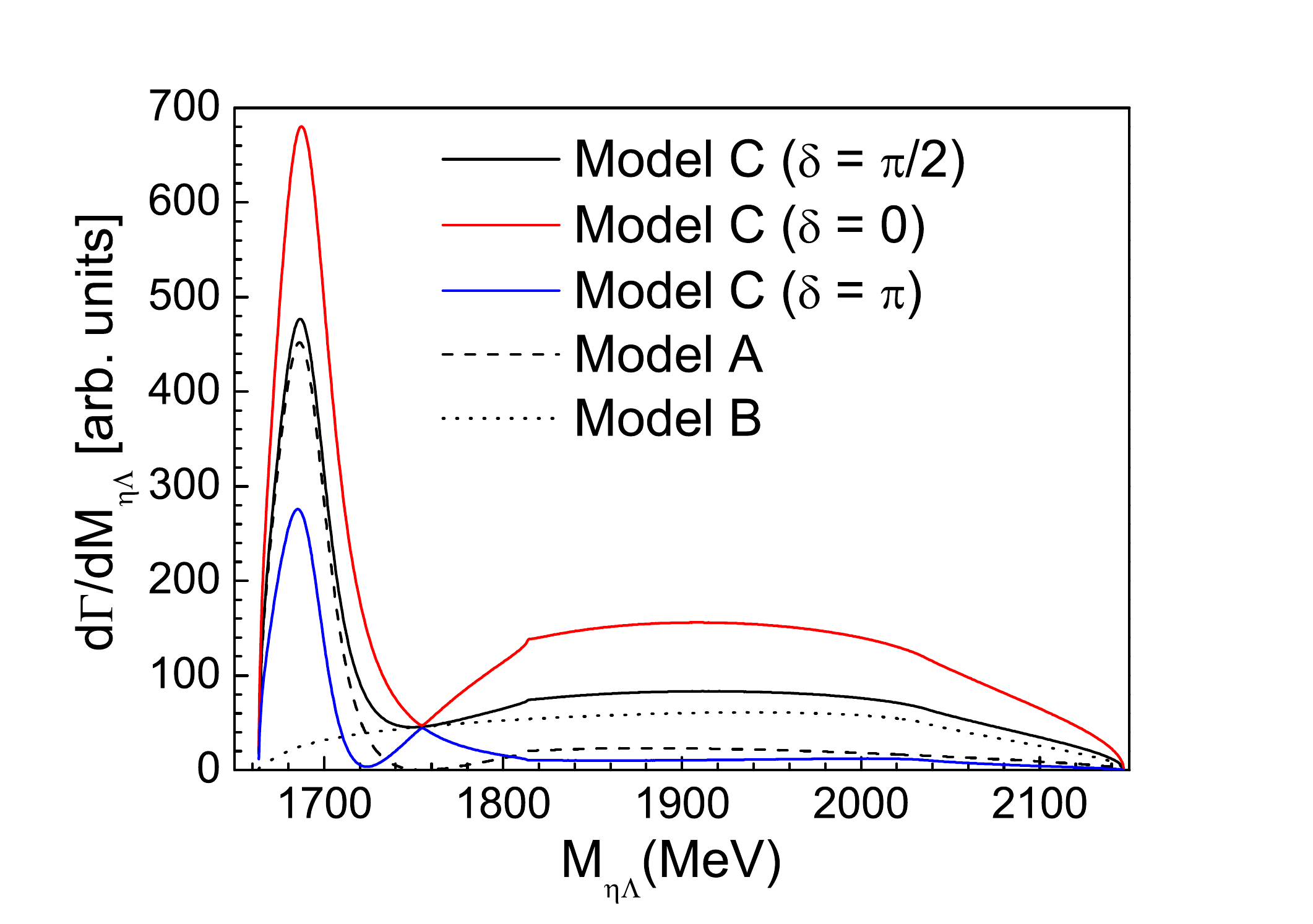}
\caption{(Color online) Invariant $\eta \Lambda$ mass distribution
for the $\Lambda^+_c \to \pi^+ \eta \Lambda$ decay. The dashed,
dotted, and solid lines represent the results obtained with Model A,
B, and C, respectively.} \label{Fig:dgdm-etalambda}
\end{center}
\end{figure}

In this section, we show the numerical results for the $\Lambda^+_c
\to \pi^+ \eta \Lambda$ decay using the formalism described in the
previous section. First we show $\frac{d\Gamma}{dM_{\eta \Lambda}}$
in Fig.~\ref{Fig:dgdm-etalambda}. The dashed and dotted curves
represent the numerical results obtained with Model A and B,
respectively, while the black-solid, red-solid, and green-solid
lines stand for the results obtained with Model C and the relative
strong phase $\delta = \pi /2$, $0$, and $\pi$, respectively. The
meson-baryon amplitudes are taken from Ref.~\cite{Oset:2001cn} and
the meson-meson amplitudes from Ref.~\cite{Oller:1997ti}. In
Fig.~\ref{Fig:dgdm-etalambda}, a peak corresponding to the $\Lambda(1670)$ resonance can be
clearly seen as in Ref.~\cite{Miyahara:2015cja}, regardless of the value of $\delta$. The interference
between $T^{MB}$ and $T^{MM}$ is instructive and destructive
 with $\delta = 0$ and $\delta = \pi$, respectively. From Fig.~\ref{Fig:dgdm-etalambda}, it is clear
that, for the $\eta \Lambda$ invariant mass distribution particularly regarding the $\Lambda(1670)$, the
effect from the meson-meson final state interaction is small and can
be safely neglected, supporting the assumption made in
Ref.~\cite{Miyahara:2015cja}. We expect that the $\Lambda(1670)$
resonance can be seen and studied from the weak decay of
$\Lambda^+_c \to \pi^+ \eta \Lambda$ in the future experiments.

\begin{figure}[htbp]
\begin{center}
\includegraphics[scale=0.45]{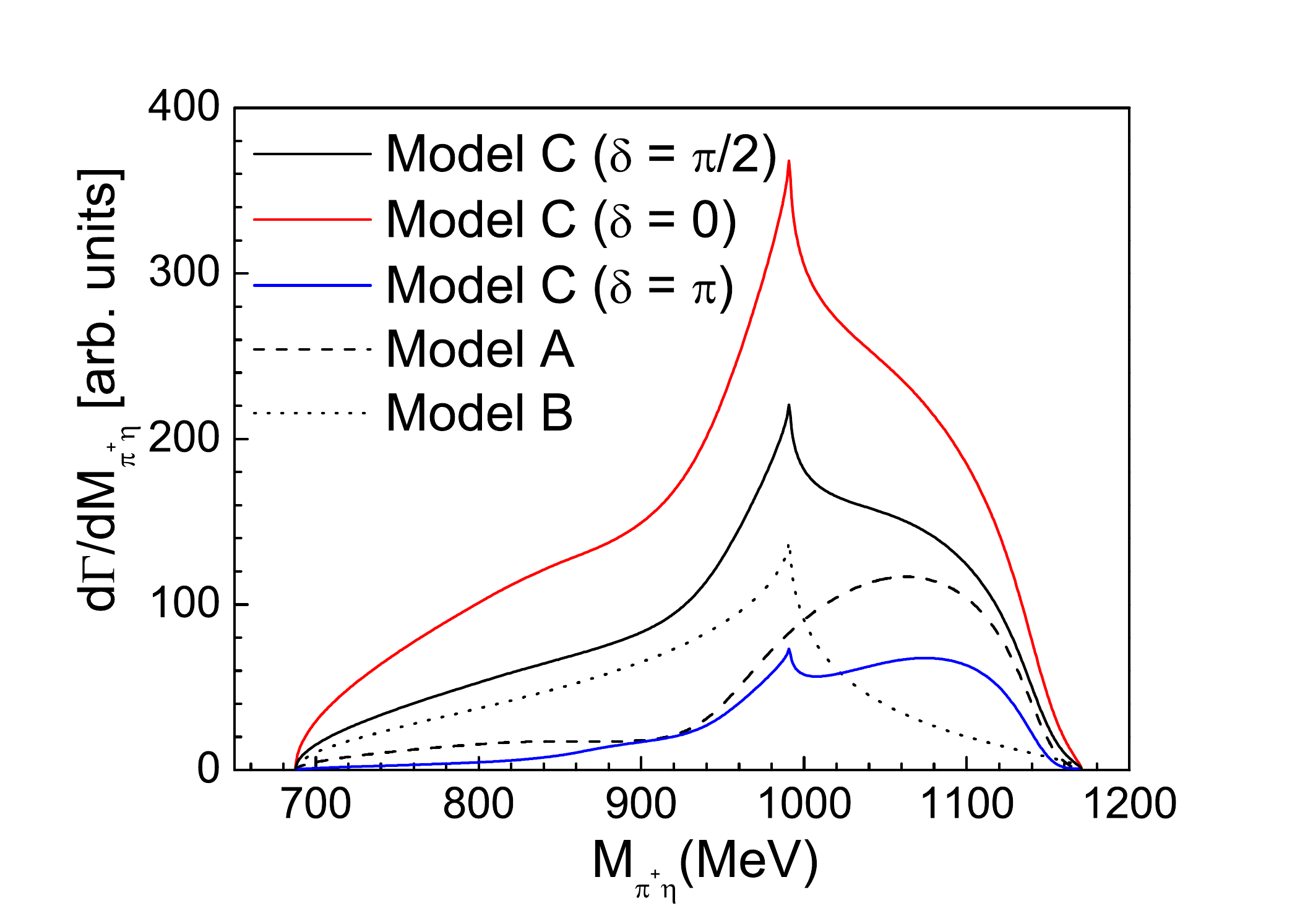}
\caption{(Color online) Invariant $\pi^+ \eta$ mass distribution for
the $\Lambda^+_c \to \pi^+ \eta \Lambda$ decay. The dashed, dotted,
and solid lines represent the results obtained with Model A, B, and
C, respectively.} \label{Fig:dgdm-pioneta}
\end{center}
\end{figure}

Next we turn to the $\pi^+ \eta$ distribution shown in
Fig.~\ref{Fig:dgdm-pioneta}. From the numerical results of Model B,
we see a clear peak/cusp around $M_{\rm inv} = 980$ MeV which
corresponds to the $a_0(980)$ state. In addition, the effect of the
meson-baryon final state interaction broadens the line shape of the
$\pi^+ \eta$ mass distribution. Nevertheless, the peak/cusp
structure of the $a_0(980)$ state is still visable. Hence,
additional experimental information on $\Lambda^+_c \to \pi^+ \eta
\Lambda$ decay can be used to investigate the nature of the
$a_0(980)$ state. It should be noted that the visibility of the
$a_0(980)$ in Model C is tied to the assumption we made for the
interference of the two hadronization amplitudes and their relative
coupling strengths. An experimental measurement of the line shape
will ultimately allow one to extract such information.

\section{Conclusions}

In the present work we have studied the $\pi^+ \eta$ and $\eta
\Lambda$ invariant mass distributions in the $\Lambda^+_c \to \pi^+
\eta \Lambda$ decay to understand better the $a_0(980)$ and
$\Lambda(1670)$ resonances. The weak interaction part is dominated
by the $c$ quark decay process: $c(ud) \to (s + u + \bar d)(ud)$,
while the hadronization part  can take place in two different
mechanisms. In the first one, the $sud$ cluster picks up a
$q\bar{q}$ pair from the vacuum and hadronizes into a meson-baryon
pair, while the $u\bar{d}$ pair from the weak decay turns into a
$\pi^+$. In the second mechanism,  the $sud$ cluster hadronizes into
a $\Lambda$, while the $u\bar{d}$ pair from the weak process
hadronizes into a meson-meson pair together with a $q\bar{q}$ pair with
the quantum numbers of the vacuum. The following final state
interactions of the meson-meson and baryon-baryon pairs are  described in
 the chiral unitary model that dynamically generates the
$a_0(980)$ and $\Lambda(1670)$ states. From the line shapes of the
invariant mass distributions, the $a_0(980)$ and
$\Lambda(1670)$ states are clearly seen.

On the experimental side, the decay mode $\Lambda^+_c \to \pi^+ \eta
\Lambda$ has been observed~\cite{Agashe:2014kda,Ammar:1995je} and
the branching ratio $\mathrm{Br}(\Lambda^+_c \to \pi^+ \eta \Lambda)
$ is determined to be $(2.4 \pm 0.5)\%$, which is one of the dominant
decay modes of the $\Lambda^+_c$ state. For the decay of
$\Lambda^+_c \to \pi^+ \eta \Lambda$, the final $\pi^+ \eta$ and
$\eta \Lambda$ states are in pure isospin $I = 1$ and $I=0$
combinations, respectively. Hence, the $\Lambda^+_c \to \pi^+ \eta
\Lambda$ decay can be an ideal process to study the $a_0(980)$ and
$\Lambda(1670)$ resonances. Future experimental measurements of the
invariant mass distributions studied in the present work will be
very helpful in illuminating  the nature of the $a_0(980)$ and
$\Lambda(1670)$ resonances. For example, a corresponding
experimental measurement could in principle be done at
BESIII~\cite{Ablikim:2015flg}.

\section{ACKNOWLEDGMENTS}

We would like to thank E.~Oset and F.~K.~Guo for useful discussions.
This work is partly supported by the National Natural Science
Foundation of China under Grant Nos. 11475227, 11375024, 11522539,
11505158, and 11475015. It is also supported by the Youth Innovation
Promotion Association CAS (No. 2016367) and the Open Project Program of State Key
Laboratory of Theoretical Physics, Institute of Theoretical Physics,
Chinese Academy of Sciences, China (No.Y5KF151CJ1).

\bibliographystyle{ursrt}

\end{document}